\documentclass[aps,twocolumn,showpacs,showkeys]{revtex4}
\usepackage{amssymb}
\usepackage{graphicx}
\newcommand{\bfm}[1]{\mbox{\boldmath${#1}$}}

\begin{document}

\title{Dissociative attachment of an electron to a molecule:\\ kinetic theory}
\author{A. Rossani}\email{alberto.rossani@polito.it} \affiliation{Istituto
Nazionale di Fisica della Materia (CNR-INFM), Dipartimento di
Fisica,\\ Politecnico di Torino, Corso Duca degli Abruzzi 24,
10129 Torino, Italy.}
\author{A.M.
Scarfone}\email{antonio.scarfone@polito.it}
\affiliation{Istituto Nazionale di Fisica della Materia
(CNR-INFM), Dipartimento di Fisica,\\ Politecnico di
Torino, Corso Duca degli Abruzzi 24, 10129 Torino, Italy.}

\date{\today}

\begin {abstract}
Test particles interact with a medium by means of a
bimolecular reversible chemical reaction. Two species are
assumed to be much more numerous so that they are
distributed according fixed distributions: Maxwellians and
Dirac's deltas. Equilibrium and its stability are
investigated in the first case. For the second case, a
system is constructed, in view of an approximate solution.
\end {abstract}

 \maketitle

\section{Introduction}

Among the elementary processes of collision between ions
and molecules, the  dissociative attachment of an electron
to a molecule
\begin{equation}
{\rm P}^-+{\rm AB}\ \rightleftharpoons\ {\rm A}^-+ {\rm PB}
\ ,\label{re}
\end{equation}
plays a role in the physics of weakly ionized gases
\cite{4,44}. However, a full kinetic study is laking. In
particular, when  the electric field is not vanishing,
equilibrium solutions are not available for the
distribution functions.\\ In the present paper we shall
assume that particles 2 and 4 are much more numerous and
can be treated as a neutral background, while particles 1
and 3 have the same (negative) charge $e$ and are subjected
to a constant electric field ${\bfm E}$. (Hereinafter,
particles P$^-$, AB, A$^-$, and PB will be labeled by the
subscripts 1, 2, 3, 4. Each of these particles is endowed
with mass $m_i$ and internal energy of chemical link
$E_i$.)

First of all we recall the full Boltzmann equations for
particles 1 and 3. The exact form of the collision
integrals is shown. The weak form of the kinetic equations
is constructed, in order to investigate conservation laws
and equilibrium. Two cases are considered:

(1) particles 2 and 4 are distributed according to
Maxwellians without drift velocity. Under this assumption,
equilibrium and its stability are investigated when ${\bfm
E}={\bfm 0}$. Firstly we show how to construct the
equilibrium solutions. Successively, we show the existence
of a Lyapunov functional for the present problem. A
physical counterpart of these mathematical results is
discussed.

(2) B is heavier than P$^-$ and A$^-$. In the limit $m_{\rm
B}\to\infty$  particles 2 and 4 can be considered to be
distributed according to Dirac's deltas \cite{3}. As a
consequence, the Boltzmann equations are modified.
Moreover, if the electric field is small, we introduce a
first order spherical harmonic expansion for the
distribution functions of particles 1 and  3. The result is
a system of differential-finite difference equations for
the problem. By taking advantage of the smallness of ${\bfm
E}$, a first order solution can be constructed for the
current functions.
\section{Boltzmann equations for particles 1 and 3}

The nonlinear integrodifferential Boltzmann equations
governing the evolution of the distribution function for
the reacting particles 1 and 3 reads as follows \cite{5}:
\begin{equation}
\left({\partial \over\partial t}+{\bfm v}\cdot{\partial
\over\partial{\bfm x}}+e\,{{\bfm E}\over
m_i}\cdot{\partial\over\partial{\bfm
v}}\right)\,f_i=J_i[{\b f}]+Q_i[{\b f}] \ ,
\end{equation}
where $J_i[{\b f}]$ and $Q_i[{\b f}]$ are the chemical and
elastic collision integrals with $\b
f\equiv(f_1,\,f_2,\,f_3,\,f_4)$.\\
The chemical collision integrals are given by
\begin{equation}
J_i[{\b f}]=\int{\cal K}_i[{\b f}]\,d{\bfm w} d{\bfm
n}^\prime \ ,
\end{equation}
where, for $i=1$, we have
\begin{eqnarray}
\nonumber &&{\cal K}_1[\b f]=
\theta(g^2-\eta_{12})\,\nu_{12}^{34}(g,\,{\bfm n}\cdot{\bfm
n}^\prime)\\
&&\times\left[\left({m_1\,m_2\over
m_3\,m_4}\right)^3\,f_3({\bfm v}_{12}^{34})\,f_4({\bfm
w}_{12}^{34})-f_1({\bfm v})f_2({\bfm w})\right],\label{k1}
\end{eqnarray}
being $\theta(x)$ the Heaviside step function whereas, for
$i=3$, we have
\begin{eqnarray}
\nonumber &&{\cal K}_3[\b f]=\nu_{34}^{12}(g,\,{\bfm
n}\cdot{\bfm n}^\prime)\\
&&\times\left[\left({m_3\,m_4\over
m_1\,m_2}\right)^3\,f_1({\bfm v}_{34}^{12})\,f_2({\bfm
w}_{34}^{12})-f_3({\bfm v})\,f_4({\bfm w})\right]
,\label{k3}
\end{eqnarray}
where
\begin{eqnarray}
&&{\bfm v}_{12}^{34}={1\over\cal M}\,(m_1\,{\bfm v}+m_2\,{\bfm w}+m_4\,g_{12}^{34}\,{\bfm n}^\prime) \ ,\\
&&{\bfm w}_{12}^{34}={1\over\cal M}\,(m_1\,{\bfm v}+m_2\,{\bfm w}-m_3\,g_{12}^{34}\,{\bfm n}^\prime) \ ,\\
&&{\bfm v}_{34}^{12}={1\over\cal M}\,(m_3\,{\bfm v}+m_4\,{\bfm w}+m_2\,g_{34}^{12}\,{\bfm n}^\prime) \ ,\\
&&{\bfm w}_{34}^{12}={1\over\cal M}\,(m_3\,{\bfm
v}+m_4\,{\bfm w}-m_1\,g_{34}^{12}\,{\bfm n}^\prime) \ ,
\end{eqnarray}
with ${\cal M}=m_1+m_2=m_3+m_4$. In Eqs. (\ref{k1}) and
(\ref{k3}) we have introduced the differential collision
frequencies of the forward and backward reaction
$\nu_{12}^{34} (g,\,{\bfm n}\cdot{\bfm n}^\prime)$ and
$\nu^{12}_{34} (g,\,{\bfm n}\cdot{\bfm n}^\prime)$, where
$g=|{\bfm v}-{\bfm w}|$ whilst ${\bfm n}$ and ${\bfm
n}^\prime$ are the unit vectors of the relative velocities
before and after collision, respectively. We observe that
the following microreversibility condition holds
\begin{eqnarray}
\nonumber & &(m_1\,m_2)^2\,g\,\nu_{12}^{34}(g,\,{\bfm
n}\cdot{\bfm
n}^\prime)=\\
&&(m_3\,m_4)^2\,g_{12}^{34}\,\nu_{34}^{12}(g,\,{\bfm
n}\cdot{\bfm n}^\prime)\, \theta(g^2-\eta_{34}) \ ,
\end{eqnarray}
where
\begin{equation}
g^{ij}_{kl}=\sqrt{{m_k\,m_l\over
m_i\,m_j}\,(g^2-\eta_{ij})} \ ,
\end{equation}
with
$\eta_{ij}=2\,{\cal M}\,\Delta E/m_i\,m_j$ and $\Delta
E=E_3+E_4-E_1-E_2>0$ is the molecular heat of reaction.

Differently, the elastic collision integrals are given by
\begin{equation}
Q_i[{\b{f}}]= \int {\cal R}_i[\b{f}]\,d{\bfm w}d{\bfm
n}^\prime \ ,
\end{equation}
where
\begin{equation}
{\cal R}_i[\b{f}]=\sum_{\ell=2,4}
\nu_{i\ell}^{i\ell}(g,\,{\bfm n}\cdot{\bfm
n}^\prime)\Big[f_i({\bfm v}_{i\ell}^{i\ell})\,f_\ell({\bfm
w}_{i\ell}^{i\ell}) -f_i({\bfm v})\,f_\ell({\bfm w})\Big] \
,
\end{equation}
and
\begin{eqnarray}
&&{\bfm v}_{i\ell}^{i\ell}={1\over m_i+m_\ell}\,(m_i\,{\bfm
v}+m_\ell\,{\bfm
w}+m_\ell\,g\,{\bfm n}^\prime) \ ,\\
&&{\bfm w}_{i\ell}^{i\ell}={1\over m_i+m_\ell}(m_i\,{\bfm
v}+m_\ell\,{\bfm w}-m_\ell\,g\,{\bfm n}^\prime) \ .
\end{eqnarray}

The weak form of the kinetic equations for $i=1$ and 3 is
obtained by multiplication times a pair of sufficiently
smooth functions $\phi_1({\bfm v})$ and $\phi_3({\bfm v})$,
respectively, integration over ${\bfm v}$, and by summing
\cite{1}:
\begin{widetext}
\begin{eqnarray}
\nonumber \int{\partial f_1\over\partial t}\phi_1({\bfm
v})\,d{\bfm v}+\int{\partial f_3 \over\partial
t}\,\phi_3({\bfm v})\,d{\bfm v}&=&\int{\cal K}_1[\b
f]\Big[\phi_1({\bfm v})-\phi_3({\bfm
v}_{12}^{34})\Big]\,d{\bfm v} d{\bfm w} d{\bfm
n}^\prime\\
&+&{1\over 2}\int\Bigg\{{\cal R}_1[\b f]\Big[\phi_1({\bfm
v})-\phi_1({\bfm v}_{12}^{12})\Big]+{\cal R}_3[\b
f]\Big[\phi_3({\bfm v})-\phi_3({\bfm
v}_{34}^{34})\Big]\Bigg\}\,d{\bfm v} d{\bfm w} d{\bfm
n}^\prime \ .
\end{eqnarray}
\end{widetext}
Observe that for $\phi_1({\bfm v})=\phi_3({\bfm v})=1$ we
get $d n_1/d t +d n_3/d t=0$, with $n_i=\int f_i\,d{\bfm
v}$, that is the total number of test particles is
conserved.

\section{Case (1)}

We assume  particles 2 and 4 much more numerous, so that
they can be treated as an equilibrium background at a fixed
temperature $T$ \cite{2}:
\begin{equation}
f_\ell=m_\ell^3\exp\Bigg[\beta\left(\mu_\ell-E_\ell-{1\over2}\,m_\ell
v^2\right)\Bigg] \ ,
\end{equation}
with $\ell=2$, 4 and $\mu_\ell$ fixed.\\ From the weak form
of the Boltzmann equation, by setting
\begin{eqnarray}
&&\phi_1=\ln\left\{\tilde f_1\,\exp\left[\beta\left(\mu_2-E_2-{1\over2}\,m_2\,v^2\right)\right]\right\} \ ,\\
&&\phi_3=\ln\left\{\tilde
f_3\,\exp\left[\beta\left(\mu_4-E_4-{1\over2}\,m_4\,v^2\right)\right]\right\}
\ ,
\end{eqnarray}
where $\tilde f_i=f_i/m^3_i$, we obtain
\begin{widetext}
\begin{eqnarray}
\nonumber {\cal D}&=&\int\nu_{12}^{34}(g,\,{\bfm
n}\cdot{\bfm n}^\prime)\,(m_1\,m_2)^3\, \ln{\tilde
f_3({\bfm v}_{12}^{34})\,\tilde f_4({\bfm
w}_{12}^{34})\over \tilde f_1({\bfm v})\, \tilde f_2({\bfm
w})}\,[\tilde f_1({\bfm v})\,\tilde f_2({\bfm w})-\tilde
f_3({\bfm v}_{12}^{34})\,\tilde f_4
({\bfm w}_{12}^{34})]\,d{\bfm v} d{\bfm w} d{\bfm n}^\prime\\
&+&{1\over 2} \sum_{i,\ell} (m_i\,m_\ell)^3\int
\nu_{i\ell}^{i\ell}(g,\,{\bfm n}\cdot{\bfm
n}^\prime)\,\ln{\tilde f_i({\bfm v}_{i\ell}^{i\ell}) \tilde
f_\ell({\bfm w}_{i\ell}^{i\ell})\over \tilde f_i({\bfm
v})\,\tilde f_\ell({\bfm w})}[\tilde f_i({\bfm v})\,\tilde
f_\ell({\bfm w})-\tilde f_i({\bfm
v}_{i\ell}^{i\ell})\,\tilde f_\ell ({\bfm
w}_{i\ell}^{i\ell})]\,d{\bfm v} d{\bfm w} d{\bfm
n}^\prime\leq0 \ ,\label{D}
\end{eqnarray}
\end{widetext}
($i=1$ and 3; $\ell=2$ and 4), where ${\cal D}$ is the left hand
side of (16). Based on these results, by
standard methods of kinetic theory \cite{5}, we have\\

{\bf Proposition 1.} The equilibrium condition
\begin{equation}
{\partial f_1\over\partial t}={\partial f_3\over \partial
t}=0 \ ,
\end{equation}
is equivalent to
\begin{eqnarray}
&&\tilde f_i({\bfm v}_{i\ell}^{i\ell})\,\tilde f_\ell({\bfm
w}_{i\ell}^{i\ell})=\tilde f_i({\bfm v})\,\tilde
f_\ell({\bfm w}) \ ,
\\ &&\tilde f_3({\bfm v}_{12}^{34})\,\tilde f_4({\bfm
w}_{12}^{34})= \tilde f_1({\bfm v})\,\tilde f_2({\bfm w}) \
,
\end{eqnarray}
with $i=1,\,3$ and $\ell=2,\,4\ \,\bullet$\\ From the first
equation we get
\begin{equation}
\tilde
f_i=\exp\left[\beta\left(\mu_i-E_i-{1\over2}\,m_i\,v^2\right)\right]
\ ,
\end{equation}
while the second one gives $\mu_1+\mu_2=\mu_3+\mu_4.$

In order to investigate the stability of such equilibrium
solution, we introduce the following functional:
\begin{widetext}
\begin{eqnarray}
{\cal L}= H &+&\beta\int
f_1\left(\mu_1-E_1-{1\over2}\,m_1\,v^2\right)\,d{\bfm
v}+\beta\int
f_3\left(\mu_3-E_3-{1\over2}\,m_3\,v^2\right)\,d{\bfm v} \
,
\end{eqnarray}
where $H=\int{\cal H}\,d{\bfm v}$, being ${\cal H}={\cal
H}_1+{\cal H}_3$ and $\partial {\cal H}_i/\partial
f_i=\ln\tilde f_i$.\\

{\bf Proposition 2.} ${\cal L}$ is a Lyapunov functional for the
present problem $\bullet$\\ {\bf Proof.} First of all, from (16)
and (20) we verify that
\begin{equation}
\frac{d\cal L}{d t}={\cal D}\,\le\,0 \ .
\end{equation}
Moreover, by introducing the first Taylor expansion of $H$
around the equilibrium
\begin{eqnarray}
\nonumber \int\hat{\cal H}\,d{\bfm v}&=&\int \left[{\cal
H}^\ast+\left(\frac{\partial{\cal H}}{\partial
f_1}\right)^*\,(f_1-f_1^*)
+\left(\frac{\partial{\cal H}}{\partial f_3}\right)^*\,(f_3-f_3^*)\right]\,d{\bfm v}\\
 &=& \int\left[{\cal H}^\ast+
\beta\left(\mu_1-E_1-{1\over2}\,m_1\,v^2\right)\,(f_1-f_1^*)
\beta\left(\mu_3-E_3-{1\over2}\,m_3\,v^2\right)\,(f_3-f_3^*)\right]\,d{\bfm
v} \ ,
\end{eqnarray}
\end{widetext}
where $\ast$ means ``at equilibrium'', from Eqs. (25) and
(27) we obtain
\begin{eqnarray}
\nonumber {\cal L}-{\cal L}^*&=&\int\Big({\cal H}-{\cal
H}^*\Big)\,d{\bfm v}\\
\nonumber&+&\beta\int\left(\mu_1-E_1-{1\over2}\,m_1\,v^2\right)\,(f_1-f_1^*)\,d{\bfm v}\\
\nonumber &+&
\beta\int\left(\mu_3-E_3-{1\over2}\,m_3\,v^2\right)\,(f_3-f_3^*)\,d{\bfm
v}\\
&=&\int({\cal H}-\hat{\cal H})\,d{\bfm v} \ .
\end{eqnarray}
Due to the convexity of ${\cal H}$ we can conclude that
${\cal L}\geq{\cal L}^*$.\\
The inequality
\begin{equation}
\frac{d{\cal L}}{dt}\leq 0 \ ,
\end{equation}
can be interpreted on a physical ground. In fact, by
introducing the entropy $S=-H$, we get the following
thermodynamic inequality:
\begin{equation}
d S\geq {1\over T}\,\left(d{\cal
E}-\mu_1^*\,dn_1-\mu_3^*\,dn_3\right) \ ,
\end{equation}
where ${\cal E}=\int f_1\,(E_1+m_1\,v^2/2)\,d{\bfm v}+\int
f_3\,(E_3+m_3\,v^2/2)\,d{\bfm v}$ is the total energy
density of test particles. With respect to Clausius
inequality, we observe an additional term due to the fact
that the medium of field particles 2 and 4 not only
provides heat to the gas of test particles 1 and 3 but also
modifies its composition.

\section{Case (2)}

In the limit $m_B\to\infty$ we have
\begin{equation}
\frac{m_1\,m_2}{m_3\,m_4}\to \frac{m_1}{m_3} \ ,
\end{equation}
and the following relations hold
\begin{eqnarray}
\nonumber &&g_{12}^{34}\to
v^-=\sqrt{\frac{m_1}{m_3}\,v^2-\eta_3} \ ,\\
\nonumber&&g_{34}^{12} \to
v^+=\sqrt{\frac{m_3}{m_1}\,v^2+\eta_1} \ ,\\ &&{\bfm
v}_{12}^{34}\to {\bfm w}+v^-{\bfm n}^\prime \ ,\hspace{5mm}
{\bfm v}_{34}^{12}\to {\bfm w}+v^+{\bfm
n}^\prime \ ,\\
\nonumber &&{\bfm v}^{i\ell}_{i\ell}\to {\bfm w}+v\,{\bfm
n}^\prime \ ,\hspace{7mm}{\bfm w}^{i\ell}_{i\ell}\to {\bfm
w}-v\,{\bfm n}^\prime \ ,\\
\nonumber&&{\bfm w}_{34}^{12}\to{\bfm w} \ ,\hspace{15.5mm}
{\bfm w}_{12}^{34}\to{\bfm w} \ ,
\end{eqnarray}
where $\eta_i=2\,\Delta E/m_i$. Moreover we can pose ${\bfm
n}\to{\bfm \Omega}$, ${\bfm
n}^\prime\to{\bfm\Omega}^\prime$ and $g\to v$.\\
\begin{widetext}
By taking into account that $f_\ell({\bfm
w})={\cal N}_\ell\,\delta({\bfm w})$  for $\ell=2$ and 4,
the integrals $J_i[\b f]$ and $Q_i[\b f]$ read now
\begin{eqnarray}
&&J_1[\b
f]=\int\theta(v^2-\eta_1)\,\nu_{12}^{34}(v,\,{\bfm\Omega}\cdot{\bfm\Omega}^\prime)\,
\left[\left(\frac{m_1}{m_3}\right)^3\,{\cal
N}_4\,f_3(v^-\,{\bfm\Omega}^\prime)
-{\cal N}_2\,f_1({\bfm v})\right]\,d{\bfm\Omega} \ ,\\
&&J_3[\b f]=\int\left(\frac{m_1}{m_3}\right)^2\,{v^+\over
v}\,\nu^{34}_{12}(v^+,\,
{\bfm\Omega}\cdot{\bfm\Omega}^\prime)
\left[\left(\frac{m_3}{m_1}\right)^2\,{\cal
N}_2\,f_1(v^+\,{\bfm\Omega}^\prime)
-{\cal N}_4\,f_3({\bfm v})\right]\,d{\bfm\Omega} \ ,\\
&&Q_1[\b f]=\int\Big[{\cal
N}_2\,\nu_{12}^{12}(v,\,{\bfm\Omega}\cdot{\bfm\Omega}^\prime)+{\cal
N}_4\,\nu_{14}^{14}(v,\,{\bfm\Omega}\cdot{\bfm\Omega}^\prime)\Big]\,
\Big[f_1(v\,{\bfm\Omega}^\prime)-f_1({\bfm v})\Big]\,d{\bfm\Omega} \ ,\\
&&Q_3[\b f]=\int\Big[{\cal
N}_2\,\nu_{32}^{32}(v,\,{\bfm\Omega}\cdot{\bfm\Omega}^\prime)+{\cal
N}_4\,\nu_{34}^{34}(v,\,{\bfm\Omega}\cdot{\bfm\Omega}^\prime)\Big]\,
\Big[f_3(v\,{\bfm\Omega}^\prime)-f_4({\bfm
v})\Big]\,d{\bfm\Omega} \ .
\end{eqnarray}
\end{widetext}
Equilibrium and its stability for the present problem are
investigated in \cite{1}. Our purpose here is to construct
model equations suitable for an approximate solution.\\
As usual in the physics of weakly ionized gases \cite{4},
if both the spatial gradients and the electric field are
small we may resort to a first order spherical harmonic
expansion of $f_i(v\,{\bfm\Omega})$:
\begin{equation}
f_i(v\,{\bfm\Omega})=N_i(v)+{\bfm\Omega}\cdot{\bfm J}_i(v)
\ ,
\end{equation}
where
\begin{equation}
N_i(v)={1\over 4\,\pi}\int
f_i(v\,{\bfm\Omega})\,d{\bfm\Omega} \ ,
\end{equation}
and
\begin{equation}
{\bfm J}_i(v)={3\over
4\,\pi}\int{\bfm\Omega}\,f_i(v\,{\bfm\Omega})\,d{\bfm\Omega}
\ .
\end{equation}
By projecting over $\bfm 1$ and $\bfm\Omega$ we get, after
some manipulations, the following system for the new
unknowns functions $F_i(\xi)=N_i(v)$ and ${\bfm
G}_i(\xi)=J_i(v)$:
\begin{widetext}
\begin{eqnarray}
\nonumber &&{\partial F_1(\xi)\over\partial
t}+{\sqrt{{\xi}}\over 3}{\bfm\nabla}\cdot{\bfm
G}_1(\xi)-{e{\bfm E}\over m_1}\cdot{2\over
3\sqrt{{\xi}}}{\partial\over\partial\xi}[\xi{\bfm
G}_1(\xi)]=\theta(\xi-\eta_1)\nu_{12(0)}^{34}(\xi)\left[\left(\frac{m_1}{m_3}\right)^3
{\cal N}_4F_3\left(\xi^-\right)-{\cal N}_2
F_1(\xi)\right] \ ,\label{f1}\\
&&\\ \nonumber &&{\partial F_3(\xi)\over\partial
t}+{\sqrt{{\xi}}\over 3}{\bfm\nabla}\cdot{\bfm G}_3(\xi)
-{e{\bfm E}\over m_3}\cdot{2\over
3\sqrt{{\xi}}}{\partial\over\partial\xi}[\xi{\bfm
G}_3(\xi)]=\left(\frac{m_1}{m_3}\right)^2
\nu_{12(0)}^{34}\left(\xi^+\right)
\sqrt{\xi^+\over\xi}\left[\left(\frac{m_3}{m_1}\right)^3{\cal
N}_2F_1\left(\xi^+\right)-{\cal N}_4F_3(\xi)\right] \
,\label{f3}
\\ &&\\ \nonumber &&{\partial{\bfm G}_1(\xi)\over\partial
t}+\sqrt{\xi}{\bfm \nabla} F_1(\xi)-\frac{2e{\bfm
E}}{m_1}\sqrt{\xi}{\partial
F_1(\xi)\over\partial\xi}=\theta(\xi-\eta_1)
\left[\nu_{12(1)}^{34}(\xi)\left(\frac{m_1}{m_3}\right)^3{\cal
N}_4{\bfm G}_3\left(\xi^-\right)-{\cal N}_2{\bfm
G}_1(\xi)\nu_{12(0)}^{34}(\xi)\right]\\&&\hspace{62mm}-\gamma_1(\xi){\bf G}_1(\xi) \ ,\\
\nonumber\\ \nonumber &&{\partial{\bfm
G_3(\xi)}\over\partial
t}+\sqrt{\xi}{\bfm\nabla}F_3(\xi)-\frac{2\,e\,{\bfm
E}}{m_3}\sqrt{\xi}{\partial
F_3(\xi)\over\partial\xi}=\left(\frac{m_1}{m_3}\right)^2\sqrt{\frac{\xi^+}{\xi}}
\left[\nu_{12(1)}^{34}\left(\xi^+
\right)\left(\frac{m_3}{m_1}\right)^3{\cal N}_2{\bfm
G}_1\left(\xi^+\right)-{\cal N}_4{\bfm
G}_3(\xi)\nu_{12(0)}^{34}\left(\xi^+\right)\right]\\
&&\hspace{63mm}-\gamma_3(\xi){\bfm G}_3(\xi) \ ,
\end{eqnarray}
\end{widetext}
where we have posed $\xi^\pm=(v^\pm)^2$ and
\begin{equation}
\gamma_i(\xi)={\cal N}_2\,\nu_{i2(t)}^{i2}(\xi)+{\cal
N}_4\,\nu_{i4(t)}^{i4}(\xi) \ ,
\end{equation}
being
$\nu_{ij(t)}^{lm}(\xi)=\nu_{ij(0)}^{lm}(\xi)-\nu_{ij(1)}^{lm}(\xi)$
and
\begin{equation}
\nu_{ij(k)}^{lm}(\xi)=2\,\pi\int_{-1}^{+1}\mu^k\,\nu_{ij}^{lm}(\xi,\mu)\,d\mu
\ ,
\end{equation}
with $k=1$ and 2.\\
Consider now the stationary space-homogeneous equations We observe
that for ${\bfm E}={\bfm 0}$ the following equilibrium solutions
hold:
\begin{equation}
F_i=C_i\,\exp\left(-{m_i\,\xi\over 2\,\kappa\, T}\right) \ ,
\end{equation}
where
\begin{equation}
{C_3\,{\cal N}_4\over C_1\,{\cal N}_2}=\left({m_3\over
m_1}\right)^3\,\exp\left(-{\Delta E\over \kappa\,T}\right) \ ,
\end{equation}
(mass action law).
\begin{figure}[ht]
\includegraphics[width=.5\textwidth]{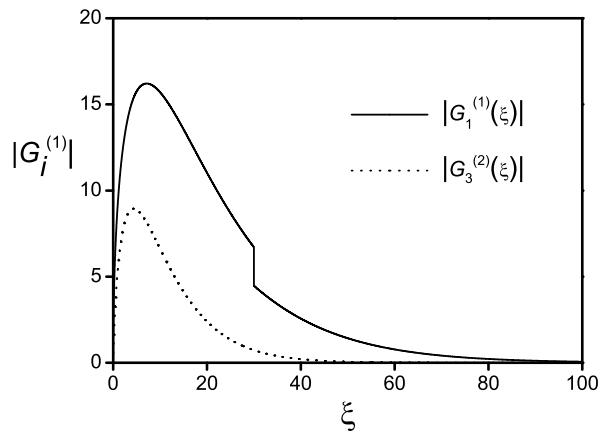}
\caption{Plot of the current function $|G_1^{(1)}(\xi)|$
(full line) for the charge particles P$^-$ and
$|G_3^{(1)}(\xi)|$ (dotted line) for the charge particles
A$^-$, in arbitrary units. }
\end{figure}
Finally, we observe that the electric field must be a small
quantity, $|{\bfm E}|=\epsilon$, so that we can expand
$F_i$ and ${\bfm G}_i$ as follows
\begin{equation}
F_i=F_i^{(0)}+\epsilon F_i^{(1)}+\ldots \ ,\hspace{10mm}
{\bfm G}_i=\epsilon\,{\bfm G}_i^{(1)}+\ldots \ .
\end{equation}
Since $F_i^{(0)} $ are already known, by solving equations
(\ref{f1}) and (\ref{f3}), we can obtain the expression of
$G_i^{(1)}$, in the case of  isotropic  reaction collision
frequency $\nu_{12(0)}^{34}$:
\begin{eqnarray}
&&G_1^{(1)}=-{e\,E\,C_1\over\kappa\,
T}\,{\sqrt{\xi}\,\exp(-m_1\,\xi/ 2\,\kappa\,
T)\over\theta(\xi-\eta_1)\,{\cal N}_2\,\nu_{12(0)}^{34}(\xi)+\gamma_1(\xi)} \ ,\label{g1}\\
\nonumber &&G_3^{(1)}=-{e\,E\,C_3\over\kappa\,
T}{\xi\,\exp(-m_3\,\xi/ 2\,\kappa\, T)\over{\cal
N}_4\left(\frac{m_1}{m_3}\right)^2\!\!\nu_{12(0)}^{34}\left(v^+
\right)v^++ \sqrt{\xi}\gamma_3(\xi)} \ ,\label{g3}\\
\end{eqnarray}
where $G_i^{(1)}={\bfm G}_i^{(1)}\cdot {\bfm e}$, with
${\bfm
e}$ the unit vector of ${\bfm E}$.\\
In figure 1 we depict, in arbitrary unity, the plots of
$|G_1^{(1)}(\xi)|$ (full line) and $|G_3^{(1)}(\xi)|$
(dotted line). Since the forward equation has a threshold
for $ \xi=\eta_1$, the collision frequency of particles 1
suddenly increases, and a discontinuity in the relevant
plot of $|G_1^{(1)}(\xi)|$ occurs.\\

\vspace{10mm}
\section{Conclusions}

Two linear Boltzmann models have been constructed for test
particles reacting with a medium of numerous field
particles. In the first case the field particles are
distributed according Maxwellians with vanishing drift
velocity. Theorems on equilibrium and its stability are
given, as well as their connection with thermodynamics. In
the second case we consider particles B heavier than
particles P$^-$ and A$^-$. By means of first order
spherical harmonic expansion, four equations can be
constructed, suitable for an approximate solution.


\begin{thebibliography}{99}

\bibitem{4} E.H. Holt and R.E. Haskell, {\em Foundations of Plasma
Dynamics}, (Macmillan, New York, 1965).

\bibitem{44} B.M. Smirnov, {\em Physics of Weakly Ionized gases}
(Mir Publishers, Moscow, 1981).

\bibitem{3} C.R. Garibotti and G. Spiga, J.
Phys. A: Math. Gen. {\bf27}, 2709 (1994).

\bibitem{5} A. Rossani and G. Spiga, Phys. A {\bf252}, 563
(1999).

\bibitem{1} M. Groppi and A. Rossani, AIP Conf. Proc. {\bf585}, 567
(2001).

\bibitem{2} M. Groppi, A. Rossani, and G. Spiga, Trans. Theor. Stat. Phys.
{\bf32}, 567 (2003).

\end{thebibliography}
\end{document}